\documentclass[a4paper,10pt]{article}
\usepackage[utf8]{inputenc}

\title{Wave Function of the Universe from a  Matrix Valued First-Order Formalism}
\author{  Sergey I. Kruglov$^1$  and Mir Faizal$^2$\\
$^1$Department of Chemical and Physical Sciences,\\ University of Toronto,
3359 Mississauga Rd. North, \\
Mississauga, Ontario L5L 1C6, Canada\\
$^2$Department of Physics and Astronomy,
\\  University of Waterloo,   Waterloo,\\
Ontario N2L 3G1, Canada
}
\date{}
\begin{document}

\maketitle

\begin{abstract}
In this paper,  the Wheeler-DeWitt equation in full superspace formalism
will be written in a matrix valued first-order formalism.
We will also  analyse  the  Wheeler-DeWitt equation
in minisuperspace approximation using this matrix valued first-order formalism.
 We will note that this  Wheeler-DeWitt equation, in this minisuperspace approximation, can be expressed as
an eigenvalue equation.
We will use this fact to analyse the spacetime foam in this   formalism. This will be done by
constructing a statistical mechanical partition function for
the  Wheeler-DeWitt equation in this matrix valued first-order formalism.
This will lead to a possible solution for the cosmological constant problem.
 \end{abstract}

\section{Introduction}

The first-order formalism in quantum mechanics first appeared in an attempt to solve the negative probabilities in the Dirac equation.
However,   with the development of second quantized quantum field theory, it was realized  that
  the first order   Dirac equation was actually the
classical field theory equation describing fermions,  and the second order  Klein-Gordon equations was
the equation describing bosons. In second quantized quantum field theory, the first quantized
wave function is interpreted as a classical field and a Lagrangian is written for it. This Lagrangian
is then used to calculate an conjugate momentum, which is in turn used for performing the second quantization
of the theory.
The second quantized wave function is a functional on the space of three-field configurations on
space-like surface.
The second quantization of a field theory can also be performed by using a functional Schroedinger's equation.
This equation for quantum gravity is called the Wheeler-DeWitt equation. It may be noted that the wave function in
the Wheeler-DeWitt equation is a functional on the space of three-metrics and matter fields projected on a space-like
surface. The wave function for the Wheeler-DeWitt equation is called the function of the universe \cite{3}-\cite{4}.
This is because in  quantum cosmology all physical information
about the universe can be extracted from this wave function \cite{1}-\cite{2c}. In other words,
the wave function of the universe describes the quantum state of the universe.
It is also possible to obtain  this  wave function   by taking a sum over all four-geometries
and field configurations, such that it matches a particular field configuration
at a spatial section of the spacetime. This method of constructing the
wave function of the universe is called the Hartle-Hawking no-boundary proposal \cite{2}.
The Wick rotation to Euclidean time makes this path integral well
defined and so this path integral approach is called Euclidean quantum gravity \cite{a}.
So, the Wheeler-DeWitt equation is a second quantized equation which can be interpreted as Schroedinger's
equation for gravity. However, the absence of time in the Wheeler-DeWitt equation occurs due to the time invariance,
which is required by general relativity \cite{5}. It may be noted that various other  boundary conditions have been analysed
for calculating the wave function of the universe. In one of the proposals the wave function of the
universe was constructed by using a quantum tunneling transition \cite{Vilenkin86}-\cite{Vilenkin82}.
In this proposal, first a baby universe is created by a quantum fluctuation of the
vacuum and then this  baby universe jumps into an inflationary period and become a parent universe. The parent
universes are the universes with a large Hubble length and baby universes are defined to be virtual fluctuations of the metric.
The existence of baby universe that give rise to a parent universe,
   naturally leads to the concept of multi-universes or multiverse.

It may be noted that just as the second quantized quantum field theory is a natural framework to study
multi-particle systems, third quantized quantum gravity is an natural framework to study multiverse \cite{third}-\cite{thir}.
This is because  in third quantization the Wheeler-DeWitt equation is viewed as a classical field equation,
and a third quantized Lagrangian for it. The functional equation of motion generated from this Lagrangian
are the Wheeler-DeWitt equation. When a third quantization is performed, the creation and annihilation
operators   create and annihilate  universes in the multiverse. In fact, it is possible to study topology change in
the multiverse by including interaction terms in the third quantized Lagrangian \cite{th1}-\cite{th2}.
It may be noted that using a model involving   baby universes    it was predicted that
the  zero cosmological constant should vanish \cite{a1}-\cite{b}.
However, we know from data obtained from type I supernovae that
our universe is an accelerating in its expansion, and so it has a small but finite  cosmological constant
\cite{super}-\cite{super5}.
There is a huge discrepancy between the predicted value of the cosmological constant and its observed value.
It is generally believed to occur from a purely quantum gravitational effect  \cite{wh}-\cite{wh1}.
It has been argued that there is no reason to assume that there is no reason to assume that the
baby universes generated from a third   quantized
Wheeler-DeWitt equations should obey Bose–Einstein statistics,
and it is possible that they obey Fermi–Dirac statistics \cite{gf}-\cite{gf1}.
It might be possible to produce a minimum value for the cosmological constant by analysing the baby
universes using this model. The first step in this direction is to construct a first-order equation from the Wheeler-DeWitt equation,
just as the Dirac equation has been constructed from the Klein-Gordon equation.
This is what we aim to do in this paper.

It may be noted that a matrix valued first-order formalism has been  used to investigate different aspects of quantum theory, that are
difficult to study using the conventional second-order formalism \cite{Kruglov}.
This formalism has been used for analysing Dirac-K\"{a}hler's massless fields \cite{k6}-\cite{6k} and for obtaining
a density matrix for the bosonic fields with two mass and spin states \cite{k4}. A first-order equation which is related to the
Proca equation in the same way as Dirac equation is related to Klein-Gordon equation is constructed using this formalism \cite{k2}.
The Podolsky generalized electrodynamics with higher derivatives has been formulated in this matrix valued first-order formalism \cite{k5}.
In fact, it has been demonstrated that the matrices of the matrix valued relativistic wave equation thus obtained obey the Petiau-Duffin-Kemmer
algebra. The solution to modified Dirac's equation with Lorentz violation terms in an external magnetic field, have been studied using this
formalism \cite{k7}. Finally, massless and spinless particles with varying speed have also been studied in this formalism \cite{k8}.
An important aspect of the first-order formalism is that it does not have the problem with negative probabilities that occur in the 
second-order formalism. The Wheeler-DeWitt equation, in minisuperspace approximation, is usually be written in the
second-order formalism. So, it also suffers from   this problem involving  negative probabilities.   
It is hoped that the first-order formalism will solve this problem. This is another reason to study this equation in the first order formalism. 
  this paper, we will use this matrix valued first-order formalism for analysing the Wheeler-DeWitt equation
in minisuperspace approximation.

\section{ Matrix Valued First-Order Formalism }

In this section, we will write the Wheeler-DeWitt equation in matrix valued first-order formalism.
The Hamiltonian constraint for general relativity  can be computed from the
Hilbert-Einstein action using the the Arnowitt-Deser-Misner decomposition \cite{adm}.
First the line element is written as
\begin{equation}
ds^{2}=g_{\mu\nu}\left(  x\right)  dx^{\mu}dx^{\nu}=\left(  -N^{2}+N_{i}%
N^{i}\right)  dt^{2}+2N_{j}dtdx^{j}+h_{ij}dx^{i}dx^{j},
\label{1}
\end{equation}
where $N_i$ is called the shift function and $N$ is called the lapse function.
Now $\sqrt{-g}R $ can be written in terms of the second
fundamental form $K_{ij}$,  and the three dimensional scalar curvature
$^3 R$ constructed from the induced metric $h_{ij}$ on a space-like surface,
\begin{equation}
\mathcal{L}\left[  N,N_{i},h_{ij}\right]  =\frac{N{}\,\sqrt{h}
}{2\kappa} \left[  K_{ij}K^{ij}-K^{2}+\,\left(  ^{3}R-2\Lambda
_{c}\right)  \right].
\label{2}
\end{equation}
where   $K$   is the
trace of the second fundamental form. The  momentum  conjugate to $h_{ij}$
can be written as
\begin{equation}
\pi^{ij}=\left( h^{ij}K-K^{ij} \right)  \frac{\sqrt{h}}{2\kappa
}.\label{mom}%
\label{3}
\end{equation}
  We calculate the Hamiltonian  using the Legendre transformation as follows:
\begin{equation}
H=
{\displaystyle\int}
 d^{3}x\left[  N\mathcal{H+}N_{i}\mathcal{H}^{i}\right]  ,
\label{4}
\end{equation}
where $
\mathcal{H}= \left(  2\kappa\right)  G_{ijkl}\pi^{ij}\pi^{kl}- {\sqrt
{h}}{}\left(  ^{3}R-2\Lambda_{c}\right)/2\kappa ,  \label{cla1}%
$ and $
\mathcal{H}^{i}=  -2\nabla_{j}\pi^{ji}.\label{cla2}%
$
Here $\Lambda_{c}$ is the cosmological constant. The equations of motion
generated from this Hamiltonian lead to two classical constraints,  $\mathcal{H} =0 $ and  $\mathcal{H}_i =0$.
The first constraint, $\mathcal{H} =0$, represents the invariance under re-parametrization and the second constraint, $\mathcal{H}_i =0$, represents the
invariance under diffeomorphism. Here $G_{ijkl}$ is given by
$
G_{ijkl}=   (h_{ik}h_{jl}+h_{il}h_{jk}-h_{ij}h_{kl})/ 2\sqrt{h}.
$
This system can be quantized by promoting
  $\mathcal{H}$ and $\mathcal{H}^{i}$ to  operators.
Thus, the quantum mechanical equation corresponding to the invariance under
diffeomorphism can be written as $\mathcal{H}_i \Psi [h] =0$, and the quantum mechanical equation corresponding to the invariance under
re-parametrization can be written as $\mathcal{H} \Psi [h] =0$. This quantum mechanical equation corresponding to the invariance under
re-parametrization is called the  Wheeler-DeWitt equation.
We could express this functional differential equation in first-order formalism, by
using a functional generalization of the matrix valued first-order formalism \cite{Kruglov}.
Thus we should in principle  look for a functional constraint of the form
$
 \Omega \Psi [h] = u (h) \Psi [h],
$
where $\Omega$ is a first-order
functional differential operator, which is functionally related the Wheeler-DeWitt equation,
in the same way as the Dirac equation is related to the Klein-Gordon  equation.
As the two indices $(i, j)$ always occur together in the Wheeler-DeWitt equation, we will represent them just by one
induce $A = (i, j)$. Now we can write the Wheeler-DeWitt equation as
\begin{equation}
 2 \kappa G_{AB}\frac{\delta }{ \delta h_{A}
 }\frac{\delta }{
\delta h_{B}} \Psi[h]  +\frac{\sqrt
{h}}{2\kappa}\left(  ^{3}R-2\Lambda_{c}\right) \Psi[h] = 0.
\end{equation}
Now we define $G_{AB} = E_{A}^a E_B^b \eta_{ab} $.
We can write the Wheeler-DeWitt equation as
 \begin{equation}
 2 \kappa E_{A}^a E_B^b \eta_{ab} \frac{\delta }{ \delta h_{A}
 }
 \frac{\delta }{
\delta h_{B}} \Psi[h]  + u^2 (h) \Psi[h] = 0,
\end{equation}
where $u^2 (h) =  {\sqrt
{h}}{}\left(  ^{3}R-2\Lambda_{c}\right)/2\kappa $.
 Now we can define $\{\Gamma_a , \Gamma_b \} = 2 \eta_{ab} I $, and write the Wheeler-DeWitt equation
 in first-order form as
  \begin{equation}
 2 i \kappa E_a^A  \Gamma^a
 \frac{D}{
D h^{A}} \Psi[h]  - u(h) \Psi[h] = 0.
\end{equation}
Here we have defined a covariant functional derivative as 
\begin{equation}
  \frac{D }{
D h_{A}}  =  \frac{\delta }{
\delta h_{A}} -   \omega^A,
\end{equation}
where  $\omega^A   = i \omega^A_{ab}[\Gamma^a, \Gamma^b ]/8$, and $\omega^A_{ab}$
is a suitable spin connection in the superspace.
It may be noted that there is no time in the Wheeler-DeWitt equation and
so the $\Gamma^a$ also  do not have a time component.
However, if we add a matter field $\phi$, then the Wheeler-DeWitt equation becomes
$
(\mathcal{H} - \mathcal{H}_m )\Psi [h, \phi] =0,
$
where $\mathcal{H}_m = \sqrt{h} T_{nn} $ is the quantum mechanical operator corresponding to the
constraint generated from a matter field.
This should also modify the form of $\Omega$ accordingly because the kinitic
part of the Wheeler-DeWitt operator will have an additional functional derivative coming from the matter fields.
It is possible to view this as time \cite{dust}-\cite{dust1}.
If we include functional derivative for the matter field as the time variable in the
Wheeler-DeWitt equation, we will be able to  follow the formalism
used   in conventional Dirac equation more closely.
However, it is difficult to proceed in this direction, and so, we will analyse
the Wheeler-DeWitt equation in first-order formalism using a minisuperspace approximation.

We will now analyse the Wheeler-DeWitt equation in a minisuperspace approximation
corresponding to a closed universe filled with a constant vacuum energy density and radiation.
The Friedman-Robertson-Walker metric for $k =1$ is given by
\begin{equation}
 ds^2 = -N^2 dt^2 + a(t) d \Omega_3^2,
 \label{6}
\end{equation}
where $d\Omega_3^2$ is the usual line element on the three sphere. The sum of the constant
energy density and radiation will be denoted by $\rho (a) = \rho_\nu + \epsilon/a^4$, where
$\rho_\nu$ is the vacuum energy density, $a$ is the scale factor and $\epsilon$ is a constant characterizing amount of radiation \cite{Michael},
\begin{equation}
 - \frac{3\pi }{4 G}a\dot{a}^2 - \frac{3 \pi}{4G}a + 2 \pi^2 a^3 \rho(a) =0.
 \label{7}
\end{equation}
It is more realistic to consider a universe filled with characterizing amount of
radiation along with the vacuum energy density.
If we assume $256 \pi^2 G^2 \rho_\nu \epsilon /9 < 1$, then a big bang occurs at $a =0$, and
the universe expands to a maximum
radius before tunneling into a phase of unbounded expansion. The Lagrangian for this system can be written as
\begin{equation}
 \mathcal{L} = - \frac{3\pi}{ 4 G} a \dot{a}^2 + \frac{3 \pi a}{4 G} - 2 \pi^2 a^3 \rho(a),
 \label{8}
\end{equation}
and so, the Hamiltonian for this system can be represented as follows:
\begin{equation}
\mathcal{H}  = - \frac{G}{3\pi}\frac{p^2}{a} - \frac{3\pi}{4G} + 2 \pi^2 a^3 \rho(a).
\label{9}
\end{equation}

The Wheeler-DeWitt equation can be obtained from constraint by promoting this Hamiltonian to an operator acting
on the wave function of the universe,
\begin{equation}
\mathcal{H}  \psi (a) = 0.
\label{10}
\end{equation}
Now we can    write the Wheeler-DeWitt equation for the universe filled with radiation and vacuum energy density as
follows 
\begin{equation}
\left[\frac{G }{3\pi}\frac{d^2}{da^2}+\frac{Gq}{3\pi a} \frac{d}{da}-\frac{3\pi}{4G}a^2+2\pi^2a^4\rho(a)\right]\psi(a)=0,
\label{11}
\end{equation}
where $ \rho(a)=\rho_v+ {\varepsilon}/{a^4}$ and $q$ is an arbitrary parameter corresponding to the
operator ordering, $p^2/a = a^{-q+1} p a^q p$.
It is convenient to represent this equation as
\begin{equation}
\left[\frac{d^2}{da^2}+\frac{q}{a}\frac{d}{da}-\frac{9 \pi^2  a^2 }{4G^2} + \frac{6\pi^3 a^4\rho(a) }{G}\right]\psi(a)=0.
\label{13}
\end{equation}
 Following the standard method used in the
matrix valued first-order formalism of quantum mechanical systems
\cite{Kruglov}, we introduce the function
\begin{equation}
 \phi = \frac{d\psi}{da}.
 \label{14}
\end{equation}
Using this function, the Wheeler-DeWitt equation can be represented in the first-order form as
\begin{equation}
\frac{d\Psi(a)}{da}+\Lambda(a)\Psi(a)=0,
\label{15}
\end{equation}
where
\begin{eqnarray}
\Psi(a)&=& \left(
\begin{array}{cc}
    \psi(a) \\
    \phi(a)
\end{array}
\right),\nonumber \\
\Lambda&=&\left(\begin{array}{cc}
  0 & -1 \\
  -\frac{9 \pi^2  a^2 }{4G^2} + \frac{6\pi^3 a^4\rho(a) }{G} & \frac{q}{a}
\end{array}\right).
\label{16}
\end{eqnarray}
It may be noted that the $2\times 2$-matrix $\Lambda$ obeys the following equation
\begin{equation}
\Lambda^2-\frac{q}{a}\Lambda- \frac{9 \pi^2  a^2 }{4G^2}+\frac{6\pi^3 a^4\rho(a) }{G} =0.
\label{17}
\end{equation}

\section{Solution to the Wheeler-DeWitt Equation}

In the previous section, we have expressed the Wheeler-DeWitt equation in a matrix valued first-order formalism.
In this section, we will solve the Wheeler-DeWitt equation in this formalism using projection matrices.
In order to do that, we first note that the Wheeler-DeWitt equation is written as an eigenvalues
equation, in this matrix valued first-order formalism. Now let  us consider the following equation eigenvalues,
\begin{eqnarray}
\frac{d\Psi(a)}{da} &=& -\Lambda\Psi(a) \nonumber \\  &=&  -\lambda\Psi(a),
\label{18}
\end{eqnarray}
where according to Caley-Hamilton theorem
$\lambda$ obeys the following  equation
\begin{equation}
\lambda^2-\frac{q}{a}\lambda - \frac{9 \pi^2  a^2 }{4G^2} + \frac{6\pi^3 a^4\rho(a) }{G} =0.
\label{19}
\end{equation}
The solutions to this equation can be written as
\begin{eqnarray}
\lambda_1&=&\frac{q}{2a}+\sqrt{\frac{q^2}{4a^2}+\frac{9 \pi^2  a^2 }{4G^2} - \frac{6\pi^3 a^4\rho(a) }{G}},
\nonumber \\ \lambda_2&=&\frac{q}{2a}-\sqrt{\frac{q^2}{4a^2}+\frac{9 \pi^2  a^2 }{4G^2} - \frac{6\pi^3 a^4\rho(a) }{G}},
\label{20}
\end{eqnarray}
so that
\begin{equation}
\left(\Lambda-\lambda_1\right)\left(\Lambda-\lambda_2\right)=0.
\label{21}
\end{equation}
Now we note that
\begin{eqnarray}
  \frac{(\Lambda-\lambda_1)^2 }{(\lambda_2-\lambda_1)^2} &=&
 \frac{\Lambda-\lambda_1}{\lambda_2-\lambda_1},  \nonumber \\
  \nonumber \\
  \frac{(\Lambda-\lambda_2)^2 }{(\lambda_1-\lambda_2)^2}&=&
 \frac{\Lambda-\lambda_2}{\lambda_1-\lambda_2}.
 \label{22}
\end{eqnarray}
We also note that
\begin{eqnarray}
  \frac{\Lambda-\lambda_1}{\lambda_2-\lambda_1} +  \frac{\Lambda-\lambda_2}{\lambda_1-\lambda_2}
 &=& 1,
 \nonumber \\
   \frac{\Lambda-\lambda_1}{\lambda_2-\lambda_1}. \frac{\Lambda-\lambda_2}{\lambda_1-\lambda_2}
  &=& 0,
  \nonumber \\
 \frac{\Lambda-\lambda_2}{\lambda_1-\lambda_2}.    \frac{\Lambda-\lambda_1}{\lambda_2-\lambda_1}
  &=& 0.
\end{eqnarray}
So, we can construct the projection matrices as \cite{Fedorov}
\begin{eqnarray}
P_1&=& \frac{\Lambda-\lambda_1}{\lambda_2-\lambda_1},\nonumber \\ P_2 &=& \frac{\Lambda-\lambda_2}{\lambda_1-\lambda_2}.
\label{24}
\end{eqnarray}
These projection matrices satisfy
$
P_1^2=P_1, \, P_2^2=P_2, \, P_1+P_2=1,$ and $ P_1P_2=P_2P_1=0.
$ Now we can  project out two  functions, $\Psi_1(a)$ and $\Psi_2(a)$ from $\Psi (a)$ as follows:
\begin{eqnarray}
\Psi_1(a)= P_1\Psi(a),
\nonumber \\
\Psi_2(a)=P_2\Psi(a).
\label{25}
\end{eqnarray}
We also have
\begin{equation}
 \Psi_1(a)+\Psi_2(a)=\Psi(a).
 \label{26}
\end{equation}
Making the transformation to another basis,
\begin{equation}
 \Psi'(a)=U\Psi(a),
 \label{27}
\end{equation}
where we define $P_1=UP_1U^{-1}$ and $P'_2=UP_2U^{-1}$, such that
\begin{eqnarray}
   ~P'_1&=& \left(\begin{array}{cc}
                                        1 & 0 \\
                                        0 & 0
                                      \end{array}\right)
                                      \nonumber \\
                 P'_2&=& \left(\begin{array}{cc}
 0 & 0 \\
 0 & 1
\end{array}\right).
\end{eqnarray}
Thus, we can write $\Lambda'=U\Lambda U^{-1}$, such that
\begin{eqnarray}
 \Lambda'&=& \left(\begin{array}{cc}
                                        \lambda_2 & 0 \\
                                        0 & \lambda_1
                                      \end{array}\right),
\nonumber \\
U&=&
                                      \left(\begin{array}{cc}
                                        \lambda_1 & 1 \\
                                        \lambda_2 & 1
                                      \end{array}\right).
\end{eqnarray}
Using this basis, we can write
\begin{eqnarray}
\Psi'_1(a)&=&\left(
\begin{array}{c}
  \psi_1(a) \\
  0
\end{array}
\right),\nonumber \\ \Psi'_2(a)&=& \left(
\begin{array}{c}
  0 \\
  \psi_2(a)
\end{array}
\right).
\label{30}
\end{eqnarray}
Now the follow equations hold
\begin{eqnarray}
\Lambda P_1&=& \lambda_2P_1,~\nonumber \\ \Lambda P_2 &=& \lambda_1P_2.
\label{31}
\end{eqnarray}
From Eq.(17) we obtain
\begin{equation}
U\frac{d\Psi(a)}{da}+U\Lambda(a)U^{-1}U\Psi(a)=0,
\label{32}
\end{equation}
Thus, we can write
\begin{equation}
\frac{d\Psi'(a)}{da}-
\mathcal{A} \psi
+\Lambda'\Psi'(a)=0,
\label{33}
\end{equation}
where
\begin{equation}
 \mathcal{A} =  \left(\begin{array}{c}
  \frac{d\lambda_1}{da} \\
 \frac{d\lambda_2}{da}
\end{array}\right).
\end{equation}
So, we obtain the following equations for the wave function of the universe,
\begin{eqnarray}
\frac{d\Psi'_1(a)}{da}-\psi\frac{d\lambda_1}{da}+\lambda_2(a)\Psi'_1(a)&=&0,\nonumber \\
\frac{d\Psi'_2(a)}{da}-\psi\frac{d\lambda_2}{da}+\lambda_1(a)\Psi'_2(a)&=&0.
\label{34}
\end{eqnarray}
Now we can write the solutions for the wave function of the universe as
\begin{eqnarray}
\Psi'_1(a)&=&\exp\left(-\int\lambda_2da\right)\Psi_{10}(a),\nonumber \\
\Psi'_2(a)&=&\exp\left(-\int\lambda_1da\right)\Psi_{20}(a),
\label{35}
\end{eqnarray}
where $\Psi_{10}(a)$, $\Psi_{20}(a)$ are two-component functions.
Thus, we have obtained the solutions to the Wheeler-DeWitt equation for a universe filled with radiation and
containing vacuum energy density in the form
\begin{equation}
\Psi'(a)=\exp\left(-\int\lambda_1da\right)\Psi_{20}(a)+\exp\left(-\int\lambda_2da\right)\Psi_{10}(a),
\label{36}
\end{equation}
where
\begin{eqnarray}
\Psi_{10}(a)&=&\left(
\begin{array}{c}
  c_2(a) \\
  0
\end{array}
\right), \nonumber \\ \Psi_{20}(a)&=& \left(
\begin{array}{c}
  0 \\
  c_1(a)
\end{array}
\right),
\label{37}
\end{eqnarray}
and $c_1(a)$, $c_2(a)$ are some functions of the scale factor $a$.
As a result, the wave function of the universe, in the first-order formalism, can be written as
\begin{equation}
\Psi'(a)=\exp\left(-\int\lambda_1da\right)\left(
\begin{array}{c}
  0 \\
  c_1(a)
\end{array}
\right)+\exp\left(-\int\lambda_2da\right)\left(
\begin{array}{c}
  c_2(a) \\
  0
\end{array}
\right),
\label{38}
\end{equation}
where
\begin{equation}
\Psi'(a)=\left(
\begin{array}{c}
  \lambda_1\psi(a)+\phi(a) \\
  \lambda_2\psi(a)+\phi(a)
\end{array}
\right),
\label{39}
\end{equation}
and as a result, we have
\begin{eqnarray}
\lambda_1\psi(a)+\phi(a)&=&c_2(a)\exp\left(-\int\lambda_2da\right),\nonumber \\
\lambda_2\psi(a)+\phi(a)&=&c_1(a)\exp\left(-\int\lambda_1da\right).
\label{40}
\end{eqnarray}
Now   we can write the solution to the second-order Wheeler-DeWitt equation using the solutions to
the first-order Wheeler-DeWitt equation,
\begin{equation}
\psi(a)=\frac{1}{\lambda_1-\lambda_2}\left[c_2(a)\exp\left(-\int\lambda_2da\right)-
c_1(a)\exp\left(-\int\lambda_1da\right)\right].
\label{41}
\end{equation}
It should be noted that the solution obtained is not singular even for $q=0$ ($a\neq 0$) because
$\lambda_1\neq \lambda_2$.  It may be noted that it has not been  not possible to write an explicit form for the functions,
$c_1 (a)$ and $c_2 (a)$.  So, now can write the wave function of the universe as
\begin{eqnarray}
 \psi(a) &=& \left[ c_2(a)\exp\left[-\int\left(\frac{q}{2a}-\sqrt{\frac{q^2}{4a^2}+\frac{9 \pi^2  a^2 }{4G^2} - \frac{6\pi^3 a^4\rho(a) }{G}}\right)da
\right] \right.  \nonumber \\  && \left.
- c_1(a)\exp\left[-\int\left(\frac{q}{2a}+\sqrt{\frac{q^2}{4a^2}+\frac{9 \pi^2  a^2 }{4G^2} - \frac{6\pi^3 a^4\rho(a) }{G}}\right)da\right] \right]
 \nonumber \\  && 
 \times \left[ 2 \sqrt{\frac{q^2}{4a^2}+\frac{9 \pi^2  a^2 }{4G^2} - \frac{6\pi^3 a^4\rho(a) }{G}}\right]^{-1}.
\end{eqnarray}

An advantage of using the first-order formalism for the Wheeler-DeWitt equation is that it is easy to analyse a 
model of spacetime foam in this formalism \cite{Hawking90}-\cite{Hawking88}. In this model the
spacetime is made up of baby universes  in which the spacetime is made baby universes at Planck scale.
Now if the baby universes are also obey the Wheeler-DeWitt equation, then we can write
the Wheeler-DeWitt equation for an ensemble of baby universes as
\begin{equation}
 \frac{d\Psi_n(a)}{da}+\Lambda_n(a)\Psi_n(a)=0.
 \end{equation}
It may be noted that an advantage of written the Wheeler-DeWitt equation in this way is that it looks like
the  Schroedinger equation in Euclidean time. So, now using the standard techniques of thermal field theory, 
we can use it  to define a temperature $T$. This is the temperature generated by the baby universes in the spacetime foam. 
These baby universes form the microstates of the same macroscopic space, 
so, we can also define a partition function for this equation as
\begin{equation}
 Z = Tr \left[\exp \left(- \frac{\Lambda_n}{T}\right)\right].
\end{equation}
Now we can write Helmholtz free energy as
\begin{equation}
 F = - T\ln Z.
\end{equation}
Now we can write the Helmholtz free energy as
\begin{equation}
 dF = -PdV -SdT.
\end{equation}
Here $P$ is the pressure generated by the baby universes.
Now the baby universes   will give space an intrinsic entropy,
which can now be written as
\begin{equation}
 S = - \left(\frac{\partial F}{\partial T }\right)_V .
\end{equation}
It may be noted that as the space has an intrinsic entropy in this approach, it will be natural for it to expand according to the
second law of thermodynamics. Furthermore, there will be an energy associated with the existence of the spacetime form. This energy
might be the reason for the existence of the cosmological constant.

\section{Conclusion}

Thus, we have obtained the solution to first-order Wheeler-DeWitt equation as well as to second-order Wheeler-DeWitt equation. It may be noted that
this model had been analysed in the WKB approximation in the second-order formalism \cite{Michael}.
In this paper, we in principle constructed a solution for this model using the matrix valued first-order formalism.
We also used the formalism for  analysing a model of   spacetime foam in which the spacetime was made up of baby universes at Planck scale.
It was noted that the existence of these baby universes can give space for an intrinsic
entropy. Furthermore, this entropy could be the reason for the existence of the cosmological constant.

The spacetime form can also be represented by a sea of
virtual black holes at Planck scale \cite{11}-\cite{1122}.
 This model has also been used as an
explanation for the final state of a real black hole. In this model,  a real  macroscopic black holes disappears in the sea of virtual
black holes after it evaporates down to the Planck size.
This model of spacetime foam also predicts that vanishing of the $\theta$-parameter which occurs in QCD.
It may be noted that these virtual black holes also
give space for an intrinsic entropy \cite{1111}. In fact, this entropy has been used as a solution for the problem of time. This is because
this entropy can be used to construct one-parameter family of unitarity operators which map any
initial density matrix to a final density matrix. Thus, this entropy naturally produces the idea of flow of time.
Higher spin field theories have also been analysed in the presence of  virtual black holes \cite{hs}.
Virtual black holes have been extensively studied in two dimensions \cite{two}-\cite{two1}.
It will be interesting to analyse virtual
black holes using a matrix valued first-order Wheeler-DeWitt equation corresponding to them.

\section*{Acknowledgement} We would like to thank Saurya Das for pointing out to us that the first-order formalism for the Wheeler-DeWitt equation 
will solve the problem with negative probabilities that occur in the second-order formalism.

\end{document}